# On the Appropriateness of Linear Stress Recovery in Biomechanical Analysis of Abdominal Aortic Aneurysm

Alastair Catlin[0009-0003-6770-9778], Mostafa Jamshidian[0000-0001-8918-5723], Adam Wittek[0000-0001-9780-8361] and Karol Miller[0000-0002-6577-2082]

Intelligent Systems for Medicine Laboratory, The University of Western Australia, Crawley WA 6009, Australia
alastair.catlin@uwa.edu.au

**Abstract.** Abdominal aortic aneurysm (AAA) wall stress is a candidate marker for rupture risk but is typically computed from single-phase images without knowledge of cardiac phase. Linear stress recovery methods, which involve solving a single geometrically linear equilibrium problem on the imaged, already-loaded geometry, have been validated for static stress estimation, but their robustness to unknown imaging phase remains unexplored. We investigated (i) whether imaging phase materially biases 99th percentile stress recovered linearly, and (ii) whether stress estimated by linear recovery agrees with stress estimated by non-linear analysis under matched loads.

Two patient-specific AAAs from a public 4D-CTA cohort (Case 1 with 5.5% measured strain and Case 2 with 4.5% measured strain) were analysed. For each case we analysed diastolic and synthetic systolic geometry, the latter generated by warping the diastolic mesh via displacement values obtained by non-linear hyperelastic analysis. Linear stresses were recovered on both geometries under systolic pressure and compared via 99th-percentile maximum principal stress values, stress percentile distributions and 3D contours showing the stress differential between the values obtained on both geometries. Additionally, linear stresses under pulse pressure were compared against non-linear stresses from the non-linear analysis and compared via 99th-percentile maximum principal stress values and stress percentile distributions.

99th-percentile stresses estimated by linear stress recovery performed on diastolic and synthetic systolic geometries under systolic pressure differed by 8.6% for Case 1 and 3.5% for Case 2 between cardiac phases, within segmentation uncertainty associated with different analysts. 99th-percentile stresses estimated by linear stress recovery and non-linear analysis under pulse pressure agreed very closely, with no difference in stress for Case 1 and just 1.1% difference for Case 2. Both cases had nearly identical distributions. These findings support linear stress recovery for patient-specific AAA analysis in clinical settings with static single-phase imaging, offering a computationally efficient alternative to non-linear methods without compromising accuracy and without the need for patient-specific mechanical properties of the AAA wall.

# 1    Introduction

Abdominal aortic aneurysm (AAA) is a lasting and irreversible dilation of the abdominal aorta that is generally asymptomatic, with diagnosis usually occurring incidentally from imaging for another condition. When untreated, an AAA can grow to the point of rupture, which in most cases, leads to death [1, 2].

Current clinical management relies primarily on maximum aneurysm diameter and growth rate, recommending intervention when diameter exceeds 5.5 cm in men or 5.0 cm in women, or when growth exceeds 1 cm/year [2]. These criteria can over- or underestimate individual rupture risk. Autopsy data indicates that 13% of AAAs ≤ 5 cm rupture, whereas 60% of AAAs > 5 cm do not [3].

AAA biomechanics, particularly wall stress estimation, has been investigated extensively for patient-specific risk assessment [4–10]. When AAA geometry is reconstructed from static medical images, such as computed tomography angiography (CTA) and magnetic resonance imaging (MRI), the segmented AAA geometry represents a loaded configuration; the stress field that equilibrates blood pressure can therefore be recovered by solving a single geometrically linear equilibrium problem with a nearly incompressible linear elastic wall, thus in effect, keeping the imaged geometry fixed [6, 7, 11, 12]. It has also been shown that residual stress effects can be taken into account via through thickness averaging consistent with Fung's Uniform Stress Hypothesis [13]. Despite the validation of linear stress recovery for recovering stress in AAAs, debate persists over the necessity of non-linear material and geometric formulations for stress recovery versus simpler linear stress recovery [4, 6].

Normally, the geometry used in a finite element analysis of an AAA is reconstructed from a 3D CT image. Information about what phase of the cardiac cycle that the 3D CT image was taken is typically not available. In this paper, taking advantage of recently acquired time resolved CTAs, we investigate how this uncertainty may affect estimated stresses. We therefore investigate: (i) whether cardiac phase at image acquisition materially alters 99[th] percentile wall stresses recovered by linear stress recovery, and (ii) how closely results from linear stress recovery and non-linear analysis agree under matched loads. We address these questions using two patient-specific AAAs which represent cases of high (5.5%) and median (4.5%) measured systolic strains from a recent 4D-CTA dataset.

# 2    Patient specific geometries and meshes

We analysed two patient-specific AAAs, whose data is available in a public 4D-CTA dataset [14]. The AAA data provided in this dataset had previously been used in two previous studies [15, 16], which investigated AAA strain and AAA structural integrity index. The two AAAs used in this study correspond to Patient 3 and Patient 7 from these two earlier studies. Patient 7, which is referred to as Case 1 in this study, was selected as it had a high measured 99[th] percentile strain of 5.5%. Patient 3, which is referred to as Case 2 in this study, was selected as it had a median 99[th] percentile strain of 4.5%.

Each of the two AAAs had two respective geometries, diastolic geometry and synthetic systolic geometry. Diastolic geometries were from the public 4D-CTA datasets, while synthetic systolic geometries were generated by warping the diastolic geometries using displacement fields obtained from non-linear finite element analysis (FEA). The non-linear FEAs were loaded with patient-specific pulse pressures of 9.5 kPa (for both Case 1 and Case 2), where pulse pressure is defined as the pressure differential of systolic and diastolic pressures and used the Raghavan-Vorp hyperelastic model [17] to represent AAA wall.

$$U = C_{10}(\bar{I}_1 - 3) + C_{20}(\bar{I}_1 - 3)^2$$

Where $U$ is the strain energy density, $\bar{I}_1$ is the first invariant of the left Cauchy–Green or Finger deformation tensor $\boldsymbol{B}$, and the constants $C_{10} = 0.174\ MPa$ and $C_{20} = 1.881\ MPa$.

We decided that this approach using synthetic systolic geometries over the true systolic geometries was more suitable for the goals of this study, as it avoided the added uncertainty associated with segmentation that using two sets of real AAA images would introduce [18]. "Intraluminal thrombus, which is often included in FEAs of AAAs, and spinal contact, were not modelled in this paper, their inclusion would not contribute to answering the two questions addressed in this paper, namely (1) whether the computed AAA wall stress is affected by the point in the cardiac cycle from which the AAA geometry for the stress analysis was captured, and (2) whether stress estimated by linear recovery agrees with stress estimated by non-linear analysis. Furthermore, our previous studies suggest mostly negligible effects of spinal contact on AAA wall stress analysis [19].

The diastolic and synthetic systolic meshes for Case 1 and 2 are shown in Figure 1 and 2 respectively. Mesh details for both cases are contained in Table 1. All meshes had three elements through the wall thickness.

**Table 1.** Mesh details for Case 1 diastolic geometry and synthetic systolic geometry and Case 2 SG and synthetic systolic geometry.

| Case | Nodes | Elements | Element Type |
|---|---|---|---|
| 1 | 796,038 | 457,042 | 10-node quadratic tetrahedrals (C3D10H in Abaqus FE code [20]) |
| 2 | 480,663 | 275,795 | 10-node quadratic tetrahedrals (C3D10H in Abaqus FE code [20]) |

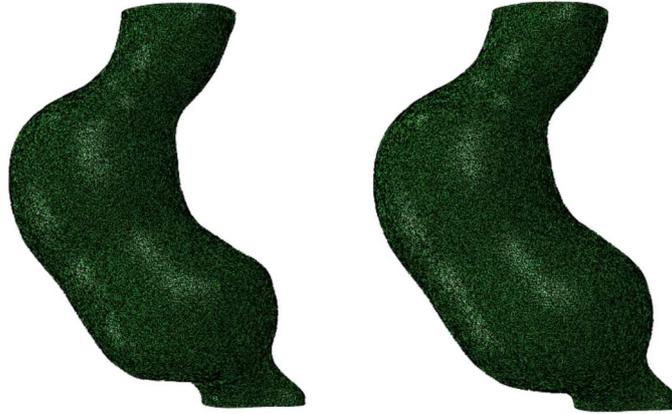

**Fig. 1** Finite element meshes of Case 1 diastolic geometry (left), and Case 1 synthetic systolic geometry (right). The difference between diastolic geometry and synthetic systolic geometry is barely perceivable (5.5% strain).

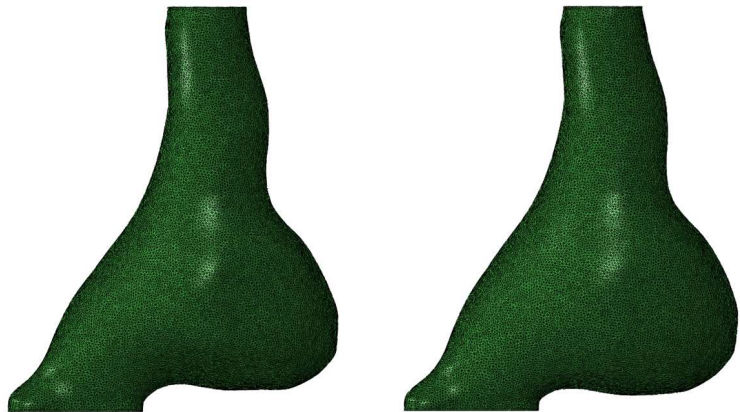

**Fig. 2** Finite element meshes of Case 2 diastolic geometry (left) and Case 2 synthetic systolic geometry (right). The difference between diastolic geometry and synthetic systolic geometry is barely perceivable (4.5% strain).

Both meshes were used in a previous study [16], which had previously validated the meshes, so no additional mesh convergence studies were conducted for these meshes.

The non-linear FEAs were conducted using Abaqus finite element code [20] on a laptop using an AMD Ryzen 7 7735U CPU and 32 GB RAM. Case 1 took 20 minutes to run whereas Case 2 took 15 minutes. 99th percentile strains obtained from the finite element analyses matched the 99th percentile strain values measured by the earlier AAA study [15].

# 3 Linear recovery of AAA stress fields at differing cardiac cycle phases

The first question investigated was whether the cardiac cycle phase at which an AAA image is captured, would significantly impact the estimated 99[th] percentile stress calculated by linear stress recovery of the AAA. To do this we conducted linear stress recovery on both the diastolic geometry and the synthetic systolic geometry for each AAA Case. Diastolic geometry and synthetic systolic geometry were both loaded with patient specific systolic pressure applied to the internal AAA walls. These two geometries represent the two extreme possibilities of where an AAA image could be captured during the cardiac cycle, with the systolic geometry being where the AAA would be at its largest geometrically, and the diastolic geometry being where the AAA would be at its smallest geometrically.

The analysis setup for the linear stress recovery of AAA performed in this paper has been described previously in the literature [6, 7, 13]. The analyses were setup by rigidly constraining both ends of the AAA, and giving the AAA wall a linear elastic wall property with a Young's modulus of 1e11 Pa and a high Poisson's ratio of 0.49, which ensures minimal compressibility, so that the FEA is essentially solving a system in static equilibrium, thus recovering stresses from the AAA geometry as segmented from the image. Furthermore, the effects of residual stress are taken into account in a post processing step by averaging the wall stresses through the wall thickness, as per Fung's Uniform Stress Hypothesis [13]. This practically eliminates the effect of wall thickness on stress magnitude. Wall stress was calculated assuming uniform wall thickness across the aneurysm. While AAA walls exhibit regional thickness variations in vivo, this assumption is mechanically justified when considering wall tension, the fundamental parameter governing vessel failure. The product of our calculated stress and the assumed uniform thickness yields wall tension (stress × thickness), which may be a more relevant mechanical quantity for rupture risk assessment [21]. Local variations in wall thickness produce inversely proportional changes in stress (i.e., thinner regions experience higher stress), such that wall tension—and consequently rupture potential—is appropriately captured despite the uniform thickness assumption [22].

## 3.1 Case 1 Results (High Strain AAA)

Figure 3 shows the maximum principal stress contours calculated for the Case 1 diastolic geometry of and for the Case 1 synthetic systolic geometry. The AAA geometries were loaded with a patient specific systolic blood pressure of 19.5 kPa.

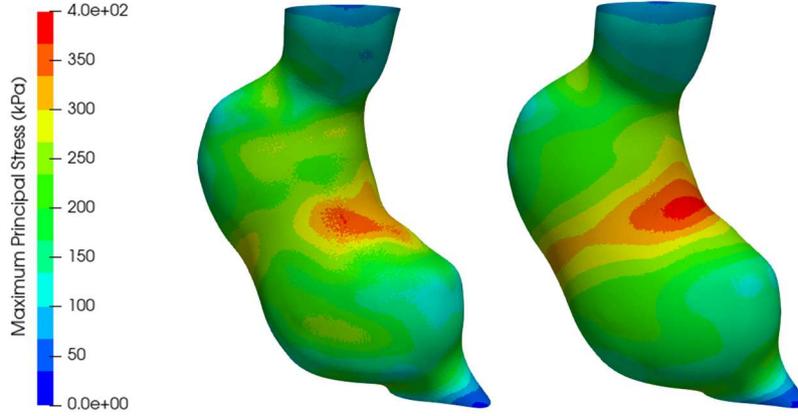

**Fig. 3** Maximum principal stress calculated using linear stress recovery on the Case 1 diastolic geometry (left) and on the Case 1 synthetic systolic geometry (right), with a patient specific systolic pressure load of 19.5 kPa. Units are kilopascals (kPa).

The 99[th] percentile maximum principal stress on the diastolic geometry was 331 kPa, which was 31 kPa, or 8.6%, lower than the 99[th] percentile maximum principal stress of 362 kPa on the synthetic systolic geometry. The difference in stresses is well within segmentation uncertainty due to inter- and intra-observer segmentation differences [18]. The stresses in the synthetic systolic geometry being higher than those in the diastolic geometry aligns with the expectation that the synthetic systolic geometry would experience greater stresses due to having a greater average diameter (46 mm) than the diastolic geometry (44 mm). The stresses in the two AAAs were predicted analytically using the equation for hoop stress in a thin-walled cylinder, which was chosen based on the geometry of the AAA more closely resembling a cylinder than another shape such as a sphere.

$$\sigma_h = \frac{pD}{2t}$$

Where $\sigma_h$ is hoop stress, $p$ is pressure, $D$ is diameter and $t$ is wall thickness.

Using a wall thickness of 1.5 mm, this equation predicts 256 kPa for the diastolic geometry, and 268 kPa for the synthetic systolic geometry, differing by factors of 1.29 and 1.35 from the 99[th] percentile stresses predicted by their corresponding FEAs respectively. The discrepancy between the stresses predicted by the FEAs and the hoop stress formula can be attributed to the geometry of the aneurysm being of an irregular shape, instead of a uniform cylindrical geometry as assumed by the equation, with areas of irregular geometry corresponding with high stress locations.

A percentile plot was produced to compare the percentile maximum principal stress distributions of both AAAs (Figure 4). The percentile plot shows that the stresses on the synthetic systolic geometry are consistently higher than the stresses on the diastolic geometry for almost all percentiles. This result is to be expected due to the larger radius of synthetic systolic geometry.

Additionally, a contour showing the interpolated nodal difference in stress between the maximum principal stresses calculated on the synthetic systolic geometry and the maximum principal stresses calculated on the diastolic geometry was produced plotted on the synthetic systolic geometry. This nodal difference contour is shown in Figure 5. The nodal difference contour shows that the maximum difference in stress is 31 kPa (8.6%), with the stress calculated on the synthetic systolic geometry being higher than those calculated on the diastolic geometry and occurring just adjacent to the peak stress location.

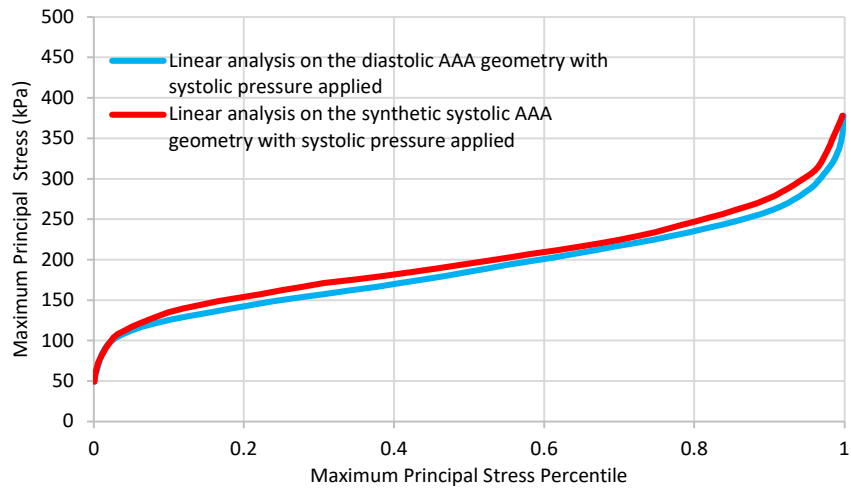

**Fig. 4** Percentile plot comparing nodal (i.e., interpolated at the nodes) maximum principal stress estimated by linear stress recovery performed on the Case 1 diastolic geometry and the Case 1 synthetic systolic geometry. Both geometries were loaded with a patient specific systolic blood pressure of 19.5 kPa.

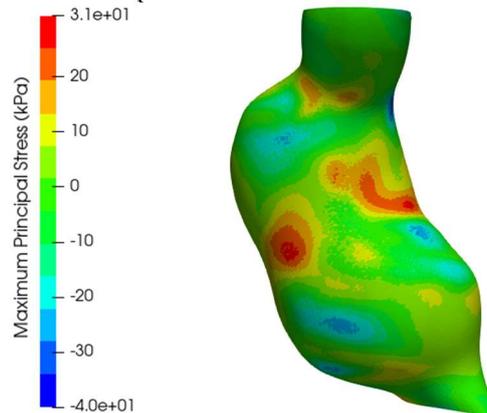

**Fig. 5** Contour showing the interpolated nodal difference in maximum principal stress calculated by linear stress recovery of the Case 1 diastolic geometry and the Case 1 synthetic systolic geometry. Both geometries were loaded with a patient specific systolic blood pressure of 19.5 kPa. Units are kilopascals (kPa).

### 3.2 Case 2 Results (Median Strain AAA)

Figure 6 shows the maximum principal stress contours calculated for the Case 2 diastolic geometry of and for the Case 2 synthetic systolic geometry. The AAA geometries were loaded with a patient specific systolic blood pressure of 19.5 kPa.

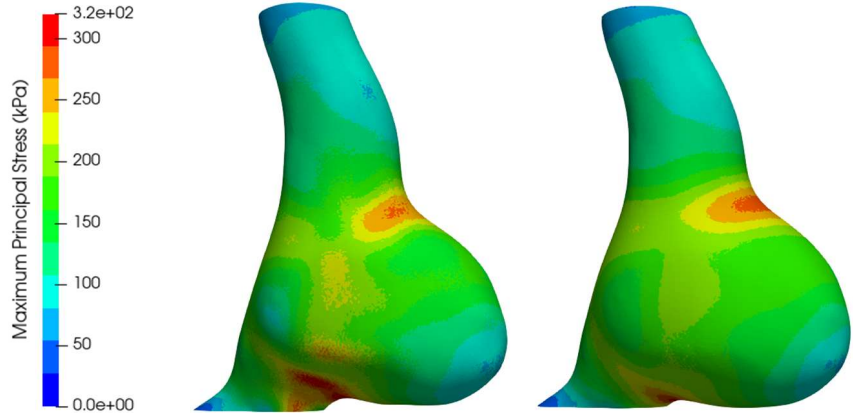

**Fig. 6** Maximum principal stress calculated using linear stress recovery on the Case 2 diastolic geometry (left) and on the Case 2 synthetic systolic geometry (right), with a patient specific systolic pressure load of 19.5 kPa. Units are kilopascals (kPa).

The 99[th] percentile maximum principal stress on the diastolic geometry was 288 kPa, which was 9.9 kPa, or 3.3%, higher than the 99th percentile maximum principal stress of 278 kPa on the synthetic systolic geometry. The difference in stresses is well within segmentation uncertainty due to inter- and intra-observer segmentation differences [18]; however, one would initially expect the synthetic systolic geometry to have higher stress than the diastolic geometry, due to the synthetic systolic geometry having a greater average diameter (52 mm) than the diastolic geometry (50.5 mm). The stresses in the two AAAs were predicted analytically using the equation for hoop stress in a thin-walled sphere, which was chosen based on the geometry of the AAA more closely resembling a sphere than another shape such as a cylinder.

$$\sigma_h = \frac{pr}{2t}$$

Where $\sigma_h$ is hoop stress, $p$ is pressure, $r$ is radius and $t$ is wall thickness.

Using a wall thickness of 1.5 mm, this equation predicts 148 kPa for the diastolic AAA, and 152 kPa for the synthetic systolic AAA, differing by factors of 1.95 and 1.83 from the 99th percentile stresses predicted by their corresponding FEAs respectively. These predicted stresses align with the stresses observed on the spherical portion of the aneurysms, however they deviate from the high stress values predicted by the finite element analysis, as the high stress values occur at regions of geometry change on the aneurysms. Additionally, the slightly higher stress in the diastolic geometry is not expected, however it can be explained by the diastolic geometry having a smaller radius of curvature at the location of maximum stress. This can be seen in Figure 8, where the synthetic systolic geometry is transparently overlaid on top of the opaque diastolic geometry. The region of maximum stress is circled in the figure.

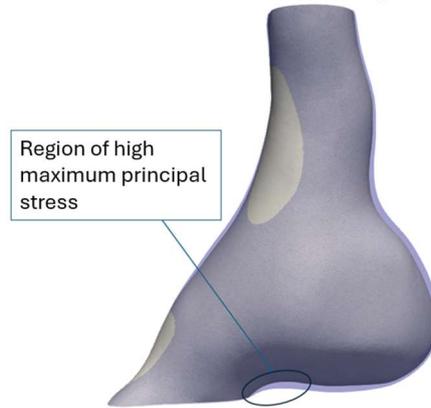

Region of high maximum principal stress

**Fig. 7** Case 2 synthetic systolic geometry overlaid transparently on top of the opaque Case 2 diastolic geometry. The region where high maximum principal stresses were calculated is circled, highlighting the difference in radius of curvature in this region between the two geometries.

A percentile plot was produced to compare the percentile maximum principal stress distributions of both AAAs (Figure 8). The percentile plot shows stresses obtained on the synthetic systolic geometry were higher than those in obtained on the diastolic geometry for most nodes until the 95th percentile, where curves align before the stresses on the diastolic geometry begin to slightly exceeds those on the synthetic systolic geometry.

Additionally, a contour showing the interpolated nodal difference in stress between the maximum principal stresses calculated on the synthetic systolic geometry and the maximum principal stresses calculated on the diastolic geometry was produced plotted on the synthetic systolic geometry. This nodal difference contour is shown in Figure 9. The contour shows a maximum difference of 9.9 kPa (3.5%) adjacent to the high stress location at the bottom of the aneurysm, with stress in the diastolic geometry being higher.

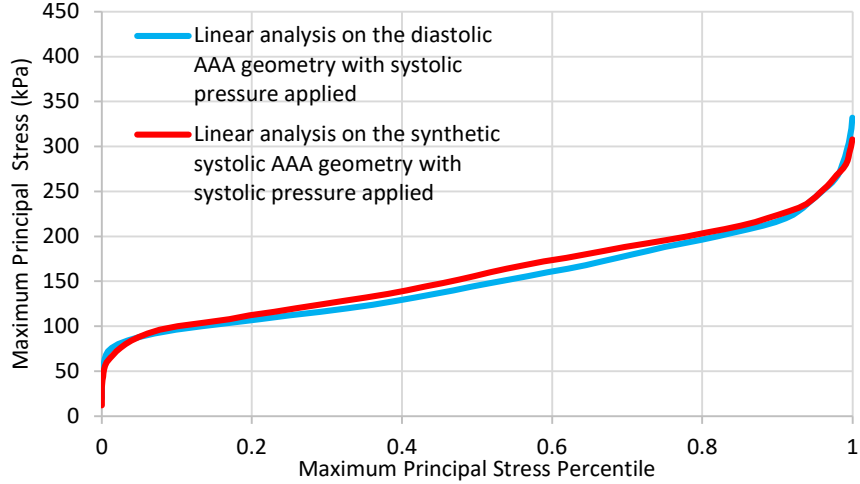

**Fig. 8** Percentile plot comparing nodal (i.e. interpolated at the nodes) maximum principal stress estimated by linear stress recovery of the Case 2 diastolic geometry and the Case 2 synthetic systolic geometry. Both geometries were loaded with a patient specific systolic blood pressure of 19.5 kPa.

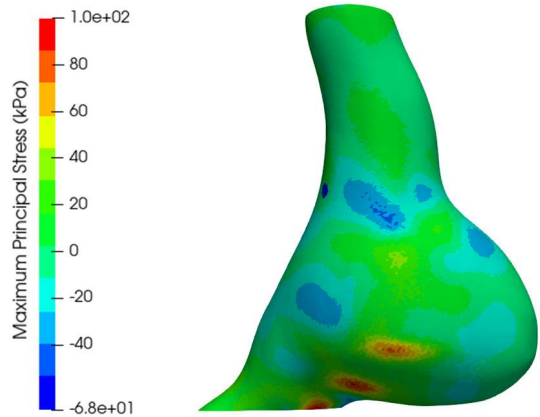

**Fig. 9** Contour showing the interpolated nodal difference in maximum principal stress calculated by linear stress recovery of the Case 2 diastolic geometry and the Case 2 synthetic systolic geometry. Both geometries are loaded with a patient specific systolic blood pressure of 19.5 kPa. Units are kilopascals (kPa).).

## 4    Non-linear and linear stress estimation of AAA stress fields

To compare the appropriateness of linear stress recovery with that of non-linear stress estimation, we calculated a third stress field using the linear stress recovery method described earlier for both AAA cases. We produced this stress field for the purpose of comparing it with the stress field produced by the non-linear analysis used to generate the synthetic systolic geometry. As such, linear stress recovery was conducted on the synthetic systolic geometry and had its internal walls loaded with a patient specific pulse pressure. The main reason that pulse pressure was used for this comparison instead of systolic pressure, is that it is the pulse pressure, the difference between the systolic and diastolic pressures, which causes a diastolic AAA to deform into its systolic state. It would not be realistic to perform a non-linear analysis on a diastolic AAA with systolic pressure applied to the inner walls, as the deformations calculated by this analysis would be unrealistically large, and not meaningful.

The results of linear stress recovery and non-linear analysis were compared via a visual comparison of stress fields and percentile plots and quantitatively by percentage difference in 99[th] percentile maximum principal stresses.

### 4.1 Case 1 Results (High Strain AAA)

Both Case 1 stress fields calculated through non-linear analysis and linear stress recovery are shown in Figure 10. The AAA geometries were loaded with a patient specific pulse blood pressure of 9.5 kPa.

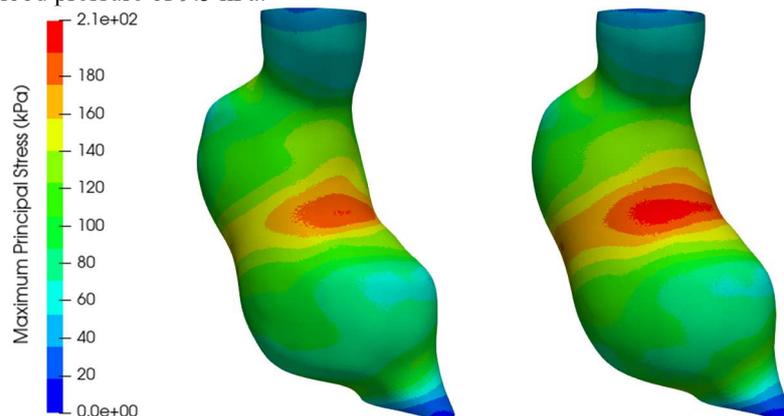

**Fig. 10** Maximum principal stress estimated on the Case 1 diastolic geometry using non-linear analysis (left) and on the Case 1 synthetic systolic geometry using linear stress recovery (right). The geometries in both analyses were loaded with a patient specific pulse pressure of 9.5 kPa. Units are kilopascals (kPa).

The 99th percentile maximum principal stresses for both analyses were 179 kPa. Percentile plots (Figure 11) were produced to compare the results of linear stress recovery and non-linear analysis and show the distributions to be almost indistinguishable.

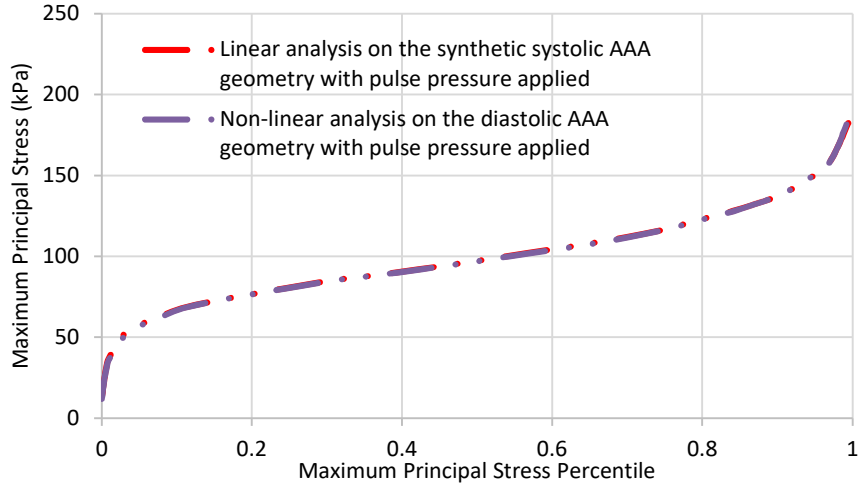

**Fig. 11** Percentile plot comparing nodal (i.e. interpolated at the nodes) maximum principal stress values estimated by the non-linear analysis of the Case 1 diastolic geometry and linear stress recovery of the Case 1 synthetic systolic geometry. The geometries in both analyses were loaded with a patient specific pulse pressure of 9.5 kPa.

### 4.2 Case 2 Results (Median Strain AAA)

Both Case 2 stress fields calculated through non-linear analysis and linear stress recovery are shown in Figure 12. The AAA geometries were loaded with a patient specific pulse blood pressure of 9.5 kPa.

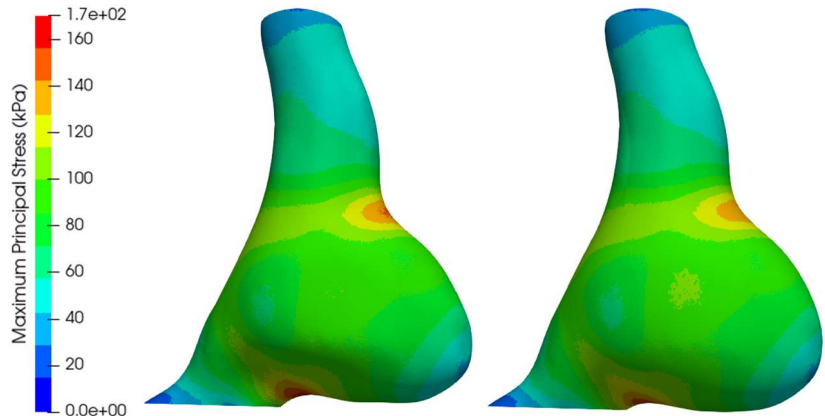

**Fig. 12** Maximum principal stress estimated on the Case 2 diastolic geometry using non-linear analysis (left) and on the Case 2 synthetic systolic geometry using linear stress recovery (right). The geometries in both analyses were loaded with a patient specific pulse pressure of 9.5 kPa. Units are kilopascals (kPa).

The 99th percentile maximum principal stresses were 144.8 kPa (non-linear) and 143.2 kPa (linear), differing by 1.6 kPa or 1.1%, well within the segmentation uncertainty due to inter- and intra-observer segmentation differences [18]. Percentile plots (Figure 13) were produced to compare the results of linear stress recovery and non-linear analysis and show the distributions to be almost indistinguishable.

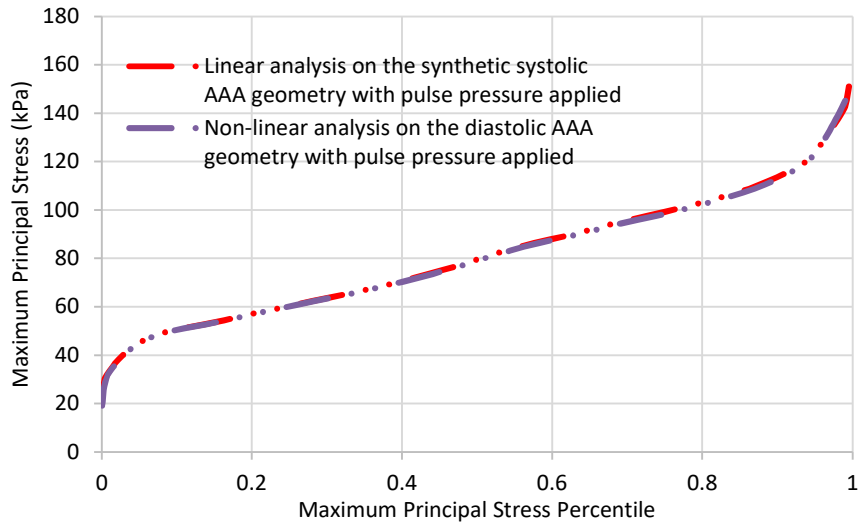

**Fig. 13** Percentile plot comparing nodal maximum principal stress values estimated by non-linear analysis of Case 2 diastolic geometry and linear stress recovery of Case 2 synthetic systolic geometry. The geometries in both analyses were loaded with a patient specific pulse pressure of 9.5 kPa.

## 5 Limitations

The prediction of AAA rupture risk and related clinical validation was beyond the scope of this study. The purpose of this study was to answer the following two questions; (i) whether cardiac phase at image acquisition materially alters 99th percentile wall stresses recovered by linear stress recovery, and (ii) how closely results from linear stress recovery and non-linear analysis agree under matched loads. For the purpose of answering these two questions, the analysis of two AAAs appears to be sufficient. Answering questions related to the clinical validity of AAA stress analysis would require a study involving many more patients and a more comprehensive AAA computational model which included ILT, however, answering such a question is beyond the scope of this study.

# 6    Conclusion

The results show that when cardiac phase is unknown, linear stress recovery methods remain appropriate for AAA biomechanical analysis. Comparing diastolic and synthetic systolic geometries under systolic pressure, 99th percentile stresses differed by 8.6% for Case 1 (331 kPa vs 362 kPa) and 3.5% for Case 2 (288 kPa vs 278 kPa), well within uncertainty due to inter- and intra-observer segmentation differences [18]. Stress in Case 2's diastolic AAA geometry being slightly higher than that in its synthetic systolic AAA geometry is explained by differing radius of curvature at peak stress locations.

It was shown that there was an even closer agreement when comparing non-linear and linear stress recovery methods, with no difference in 99th percentile stress for Case 1 (both 179 kPa) and 1.1% difference in 99th percentile stress for Case 2 (144.8 kPa vs 143.2 kPa), with nearly identical percentile distributions across all stress levels. These findings support the use of linear stress recovery in patient-specific AAA analysis in clinical settings when only static single-phase imaging is available, offering a computationally efficient alternative to non-linear methods without compromising accuracy and without the need for patient-specific mechanical properties of the AAA wall.

# 7    Acknowledgements

This work was supported by the Australian National Health and Medical Research Council NHMRC Ideas grant no. APP2001689. This study uses images from public 4D-CTA dataset [14]. Contributions of Christopher Wood and Jane Polce, radiology technicians at the Medical Imaging Department, Fiona Stanley Hospital (Murdoch, Western Australia) to acquisition of these images at Fiona Stanley Hospital, and Dr Farah Alkhatib's (The University of Western Australia) assistance in the image acquisition are gratefully acknowledged. When conducting this study, A.C. was supported by an Australian Government Research Training Program (RTP) Scholarship.